\documentclass[12pt]{iopart}
\usepackage{iopams,graphicx,braket,color}  
\usepackage[utf8]{inputenc}
\begin{document}

\title{Phases of cold atoms interacting via photon-mediated long-range forces}
\author{Tim Keller$^1$ , Simon B. J\"ager$^1$, and Giovanna Morigi$^1$}

\address{$^1$ Theoretische Physik, Saarland University, D-66123 Saarbruecken, Germany}

\begin{abstract}
Atoms in high-finesse optical resonators interact via the photons they multiply scatter into the cavity modes. The dynamics is characterized by dispersive and dissipative optomechanical long-range forces, which are mediated by the cavity photons, and exhibits a steady state for certain parameter regimes. In standing-wave cavities the atoms can form stable spatial gratings. Moreover, their asymptotic distribution is a Maxwell-Boltzmann whose effective temperature is controlled by the laser parameters. In this work we show that in a two-mode standing-wave cavity the stationary state possesses the same properties and phases of the Generalized Hamiltonian Mean Field model in the canonical ensemble. This model has three equilibrium phases: a paramagnetic, a nematic, and a ferromagnetic one, which here correspond to different spatial orders of the atomic gas and can be detected by means of the light emitted by the cavities. We further discuss perspectives for investigating in this setup the ensemble inequivalence predicted for the Generalized Hamiltonian Mean Field model.
\end{abstract}

\date{\today}

\maketitle

%
%
%
\maketitle
%

\section{Introduction}

Atomic ensembles in optical resonators offer a promising platform for studying the physics of long-range interacting systems \cite{Bachelard:2010}. The long-range interaction here originates from multiple scattering of cavity photons, which carry the information about the positions of the scattering atoms and thus mediate an optomechanical interparticle potential  \cite{Ritsch:2013}. In a single-mode cavity the photons are coherent over the cavity mode volume, which makes the interaction range as large as the system size. Therefore, the energy is non-additive like in gravitational and Coulomb systems in two or more dimensions  \cite{Bachelard:2010,Schuetz:2014,Campa:2009}. 

In equilibrium statistical mechanics, consequences of non-additivity are for instance the super-linear scaling of thermodynamic quantities with the system size and the inequivalence of the statistical ensembles \cite{Campa:2009}, one manifestation of which are negative specific heats in the microcanonical ensemble  \cite{Campa:2009,Hertel:1971,Barre:2001}. Differing from these systems, however, the dynamics of atomic gases in optical cavities is typically dissipative and non-trivial effects can only be observed if either the atoms or the cavity are pumped by light \cite{Ritsch:2013,Schuetz:2016}. The steady state, when it exists, results from the dynamical interplay between drive and losses and its properties thus depend on the drive and on the cavity parameters. It is therefore often not possible to draw a direct connection with equilibrium statistical mechanics of long-range interacting systems. 

In this context it is remarkable that  for some parameter regimes the dynamics of atoms' spatial selforganization in an optical resonator can be mapped to long-range interacting systems at equilibrium \cite{Ritsch:2013,Domokos:2002, Black:2003}. Selforganization of the atomic gas in ordered spatial patterns occurs in a single-mode standing-wave resonator when the atoms are driven by lasers whose intensity exceeds a threshold value, which depends also on the cavity decay rate \cite{Ritsch:2013,Domokos:2002, Black:2003,Asboth:2005,Schuetz:2015}. By suitably tuning the laser frequency, moreover, a stationary state exists which is characterised by a Maxwell-Boltzmann distribution of the atomic momentum. In Ref. \cite{Schuetz:2015} it was shown that the stationary dynamics can be mapped to the one of the Hamiltonian-Mean-Field model in a canonical ensemble \cite{Campa:2009,Antoni:1995} and in particular that the transition to spatial order can be described in terms of a Landau second-order phase transition. The dynamics leading to equilibrium, moreover, exhibits a slow relaxation that is due to the interplay between the conservative and dissipative cavity-mediated long-range forces \cite{Schuetz:2016}.

\begin{figure}
\includegraphics[width=\textwidth]{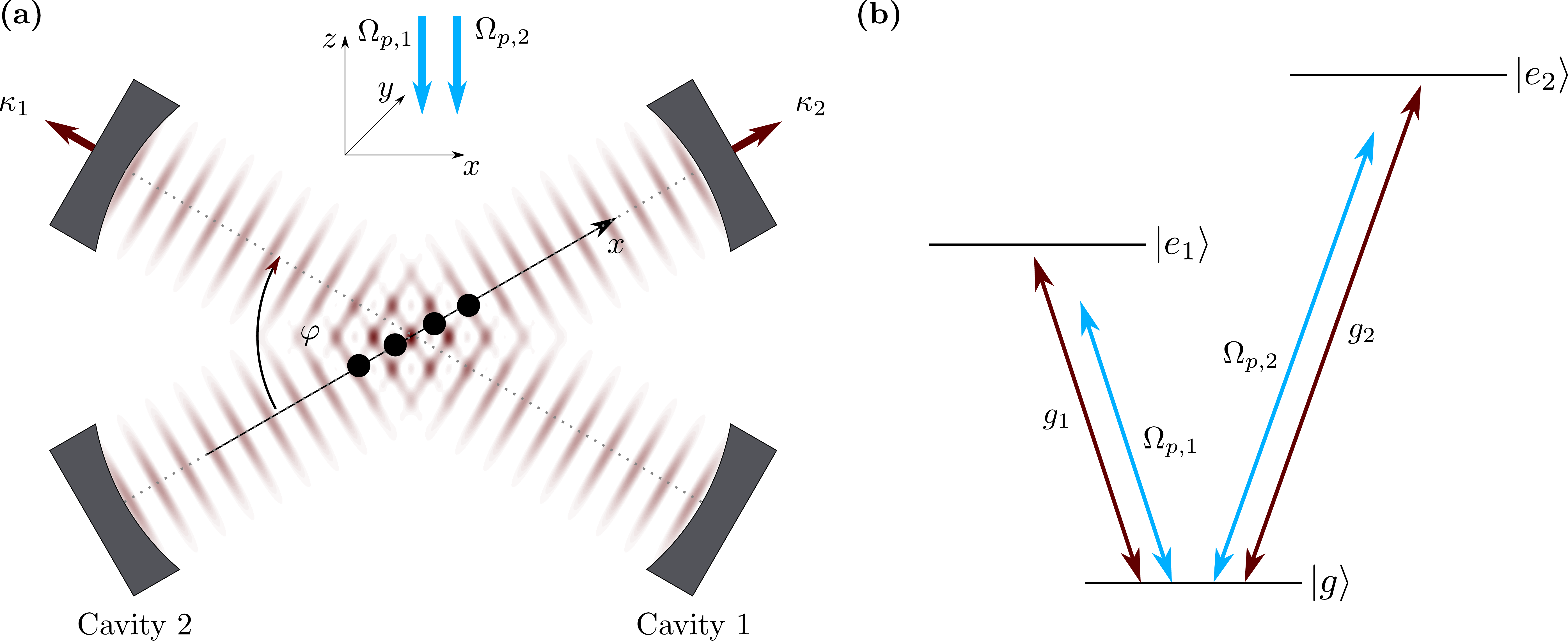}
\caption{\textbf{(a)} A gas of cold atoms is confined to move along the $x$-axis and interacts with the modes of two optical cavities, whose wave vectors form the angle $\varphi$. The cavities emit photons at rate $\kappa_1$ and $\kappa_2$, respectively, and are pumped via coherent scattering of laser photons by the atoms. The atoms, in turn,  experience the optical potential and the dissipative forces which result from the mechanical effects of the two cavity fields. Inset: Linearly-polarized lasers propagate along the direction orthogonal to the plane defined by the two cavity wave vectors. We assume that cavity 1(2) couples with the electronic transition $|g\rangle\to |e_1\rangle$ ($|g\rangle\to |e_2\rangle$), as illustrated in subplot \textbf{(b)}. The electronic transitions are also driven by the laser fields at Rabi frequency $\Omega_{p,j}$ ($j=1,2$). The cavity and laser fields are far-off resonance from the dipolar transition they couple to and quasi resonant with each other.  In this limit the scattering is prevailingly coherent. 
} 
\label{Fig:1}
\end{figure}

In this paper we extend the model of Ref. \cite{Schuetz:2015} and consider a gas of cold atoms that interact with two cavity modes and are transversally driven by lasers. A possible setup is illustrated in Fig. \ref{Fig:1}. We determine the parameter regimes where the dynamics asymptotically tends to a stationary state and show that  its phase diagram as a function of the lasers' and of the cavity parameters can be mapped to the one of the Generalized Hamiltonian Mean Field model (GHMF) in a canonical ensemble \cite{Teles:2012,Levin:2014,Pikovsky:2014}. This model describes the dynamics of $N$ particles with canonically conjugated variables $p_j$, $\theta_j$ constrained on a circle that interact via competing long-range forces. In the form studied in Ref. \cite{Teles:2012,Pikovsky:2014} its Hamiltonian reads
\begin{equation}
\label{eq:GHMF}
H=\sum_j\frac{p_j^2}{2}+\frac{1}{2N}\sum_{i,j}\left(1-\Delta\cos\theta_{ij}-(1-\Delta)\cos2\theta_{ij}\right)\,,\label{GHMF}
\end{equation}
where $\theta_{ij}=\theta_i-\theta_j\in[0,2\pi)$ and $\Delta$ is a dimensionless parameter that can vary continuously in the interval $[0,1]$. The phase diagram as a function of the temperature and $\Delta$ is characterised by (i) a paramagnetic, (ii) a nematic, and (iii) a ferromagnetic phase, with first and second-order transitions. In our case the phases correspond to density modulations of the atoms at different periodicity and can be detected through the light emitted by the cavity. Our motivation draws from ongoing experimental investigations \cite{Leonard:2016}. Theoretical studies of this system focused on the dynamics leading to equilibrium and reported the existence of several metastable states \cite{Kraemer:2014}. These properties are at the basis of proposals for using these systems to simulate a quantum Hopfield associative memory scheme \cite{Torggler:2014,Torggler:2016}. The determination of the condition for a stationary state of the setup in Fig. \ref{Fig:1} and the analysis of the corresponding phase diagram is the main result of the present manuscript. The mapping to the GHMF model shows that Cavity Quantum Electrodynamics (CQED) setups offer a versatile platform for studying the statistical mechanics of systems with long-range interactions. 

This article is organised as follows. In Sec. \ref{Sec:2} we introduce the physical model and sketch the derivation of a Fokker-Planck equation governing the dynamics of the atoms' external variables in the semiclassical limit. We then determine the parameters' regime for which the Fokker-Planck equation allows for a stationary state which is a Maxwell-Boltzmann distribution. In Sec. \ref{Sec:3} we define the free energy, which we can associate to the stationary state, and introduce an appropriate thermodynamic limit. We then show that the free energy can be mapped to the one of the GHMF model. We determine the phase diagram as a function of the system's parameters and identify the observables which allow one to measure the predicted phases. In Sec. \ref{Sec:4} we discuss possible implementations of the setup which could correspond to the realization of the microcanonical ensemble of the GHMF model, for which ensemble inequivalence has been predicted \cite{Pikovsky:2014}, and identify the parameters regimes for which it could be measured. 

\section{Semiclassical dynamics of an atomic gas in an optical cavity}\label{Sec:2}

The system we consider consists of a gas of $N$ cold atoms of mass $m$, whose motion is confined to occur along one dimension parallel to the unit vector ${\bf e}_x$. We denote by $\hat x_j$ and $\hat p_j$ the canonically conjugated position and momentum operators ($j=1,\ldots,N$), such that $\left[\hat{x}_i,\hat{p}_j\right]=i\hbar\delta_{ij}$. The atoms experience the optomechanical forces due to the interaction with the lasers and with the quantized fields of two high-finesse cavities in the setup of Fig. \ref{Fig:1}. Specifically, the axes of the two cavities are in the $x-y$ plane, the wave vector ${\bf k}_1$ of cavity 1 forms an angle $\varphi$ with the $x$-axis and thus the force the atoms experience is the projection of the mechanical force along $x$, while the wave vector ${\bf k}_2$ of cavity 2 is parallel to ${\bf e}_x$. The laser fields are linearly polarised and propagate in the direction orthogonal to the plane, they pump the cavity fields by means of coherent scattering via the atoms. The amplitude of coherent scattering, in turn, is maximal when the atoms form Bragg gratings, whose stability depends on the mechanical forces of the cavity light. As we will show, a nematic phase corresponds to a stable Bragg grating which supports the build-up of the field of only one cavity mode. In the ferromagnetic phase, instead, the atoms form stable Bragg gratings for both modes.

Below we describe the setup in detail and introduce the master equation for the density matrix $\hat{\rho}$ of atoms and cavity fields which governs the system's dynamics. We then sketch the derivation of a Fokker-Planck equation for the motion of the atoms' external degrees of freedom, which is valid when the atomic variables can be treated as semiclassical variables and the cavity fields can be eliminated from the equations of motion in a coarse-grained time scale. We finally determine the stationary state of the atoms and identify the regime in which it is a thermal state.

\subsection{Master equation}

The state of the atoms' external degrees of freedom and of the cavity modes is described by the density operator $\hat{\rho}$, whose dynamics is governed by a Born-Markov master equation of the form: 
\begin{equation}
\frac{\partial}{\partial t}\hat{\rho}=\frac{1}{i\hbar}\left[\hat{H}_0,\hat{\rho}\right]-\sum_{n=1,2}\kappa_n\left(\hat{a}^\dagger_n\hat{a}_n\hat{\rho}+\hat{\rho}\hat{a}^\dagger_n\hat{a}_n-2\hat{a}_n\hat{\rho}\hat{a}^\dagger_n\right)\,,
\label{eq:master_equation}
\end{equation}
where $\hat H_0$ is the Hamiltonian of the system, which we introduce below, and the second term of the right-hand side describes photon emission by the cavity modes at rate  $\kappa_n$ ($n=1,2$). Here, $\hat{a}^\dagger_n$ and $\hat{a}_n$ denote the creation and annihilation operators of a photon of the standing-wave cavity mode $n$ ($n=1,2$), with wave vector ${\bf k}_n$, frequency $\omega_{c,n}=c|{{\bf k}_n}|$,  and linear polarization in the $x-y$ plane. The commutation relations are $\left[\hat{a}_i,\hat{a}^\dagger_j\right]=\delta_{ij}$. 

Hamiltonian $\hat H_0$ describes the optomechanical coupling between atoms' degrees of freedom and cavity modes. It is an effective Hamiltonian derived in the limit where the atoms' internal degrees of freedom can be adiabatically eliminated, such that the scattering processes are coherent and the relevant parameters of the atomic internal structure is the atoms' polarizability \cite{Vukics:2005}. Here, cavity mode $n$ couples with the electronic transition $|g\rangle\to |e_n\rangle$ at frequency $\omega_{a,n}$ with vacuum Rabi frequency $g_n$. We further assume that the coupling of mode 1 (2) with $|g\rangle\to |e_{2}\rangle$ ($|g\rangle\to |e_{1}\rangle$) is off-resonance by orders of magnitude and can be discarded (nonetheless, the wave numbers are assumed to be $|{\bf k}_1|\approx |{\bf k}_2|=k$). In this limit cavity $n$ is pumped by coherent scattering of the laser, which couples to the transition $|g\rangle\to |e_n\rangle$ with Rabi frequency $\Omega_{p,n}$ and frequency $\omega_{p,n}$. The condition for adiabatic elimination of the internal excited state is given by the inequality $|\omega_{a,n}-\omega_{p,n}|,|\omega_{a,n}-\omega_{c,n}|\gg \Omega_{p,n},g_n\sqrt{\bar n_{c,n}},|\Delta_n|$, where $\bar n_{c,n}$ is the mean intracavity photon number in cavity $n$ and $\Delta_n=\omega_{p,n}-\omega_{c,n}$ is the detuning of the laser from the cavity mode it pumps \cite{Schuetz:2014}. In this regime $\hat H_0$ reads:
\begin{eqnarray}
\hat{H}_0=\sum_{i=1}^{N}\frac{\hat{p}_i^2}{2m}&-\hbar\sum_{n=1,2}\left(\Delta_n-U_n\sum_{i=1}^{N}\cos^2(k_n\hat{x}_i)\right)\hat{a}^\dagger_n\hat{a}_n\nonumber\\
&+\hbar\sum_{n=1,2}S_n\sum_{i=1}^{N}\cos(k_n\hat{x}_i)\left(\hat{a}^\dagger_n+\hat{a}_n\right)\,,
\label{eq:effective_hamiltonian}
\end{eqnarray}
and is here reported in the frame where each atomic transition and cavity mode rotates at the corresponding laser frequency. Beside the kinetic energy of the atoms, it contains the resonators' energy, which is shifted by the dynamical Stark shift with amplitude $U_n$ induced by the coupling between cavity mode and the atoms at the position $x_i$ within the cavity spatial mode function $\cos(k_n\hat x_i)$. This term is also a periodic potential for the atoms whose depth is a dynamical variable. The last term on the right-hand side, finally, describes coherent scattering by the atoms between laser and cavity mode with coupling strength $S_n=g_n\Omega_{p,n}/(\omega_{p,n}-\omega_{a,n})$. It is an effective pump of the resonator whose amplitude is maximal when the atoms form Bragg gratings maximizing the expectation value of the operator $\sum_{i=1}^{N}\cos(k_n\hat x_i)$. 

Note that in Eq. (\ref{eq:effective_hamiltonian}) we introduced the notation $k_1\equiv|{\bf k}_1\cdot {\bf e}_x|=k\cos\varphi$ and $k_2\equiv|{\bf k}_2\cdot{\bf e}_x|=k$. In the following we will set $\varphi=\pi/3$, thus $k_1=k/2$.

\subsection{Fokker-Planck equation for the atoms' external variables}

\begin{figure}[h!]
\center\includegraphics[width=1\textwidth]{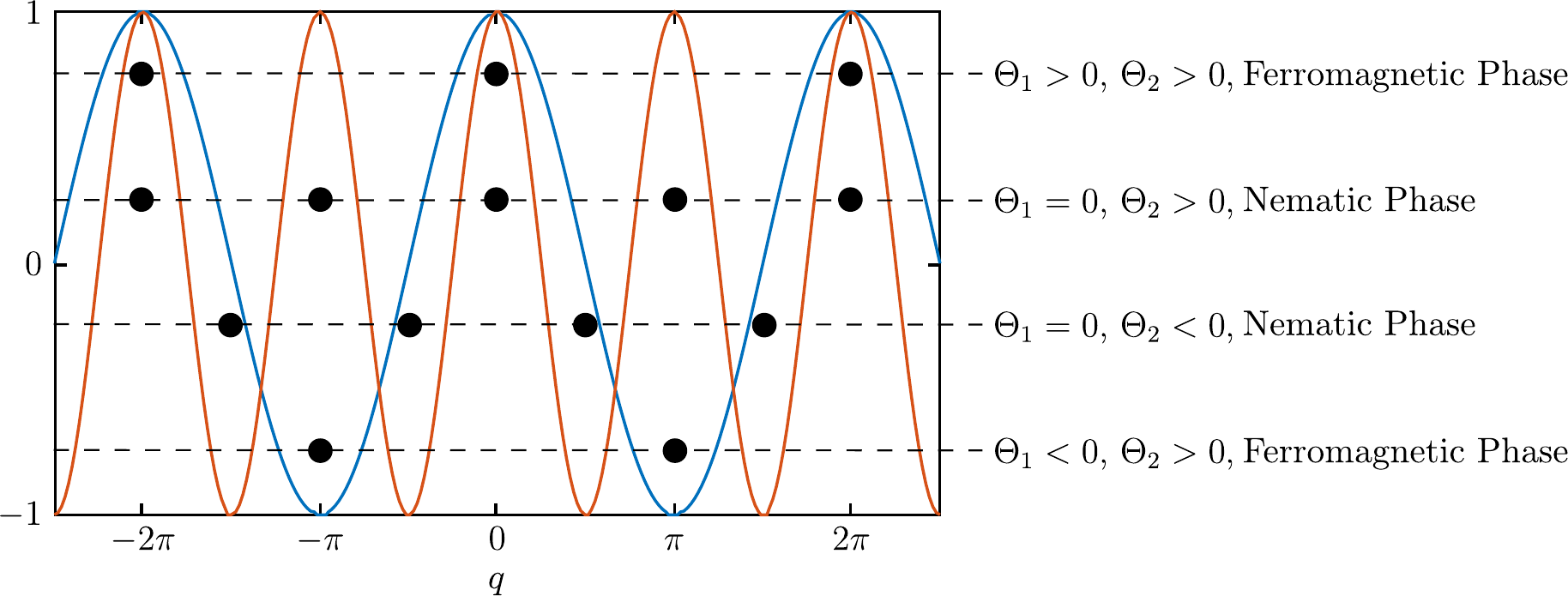}
\caption{Periodic potential of cavity 1 (blue line) and cavity 2 (red line) along the $x$-axis and as a function of $q=kx/2$. The spatial configurations leading to non-vanishing values of $\Theta_1$ and/or $\Theta_2$ are illustrated by the bullet points along each horizontal line. As we show below, configurations where $\Theta_1=0$ while $\Theta_2\neq 0$ correspond to the nematic phase of the GHMF. Configurations where both $\Theta_1,\Theta_2\neq 0$ are ferromagnetic phases. Each configuration gives rise to different cavity field amplitudes and can thus be detected by monitoring the fields at the cavities' outputs.} 
\label{Fig:2}
\end{figure}

We now discuss the assumptions at the basis of the derivation of a Fokker-Planck equation (FPE) for the dynamics of the atomic external variables. A semiclassical description of the atoms' center-of-mass motion is justified when the width $\Delta p$ of the single-atom momentum distribution is much larger than the linear momentum $\hbar k$ carried by a cavity photon $\Delta p \gg\hbar k$ \cite{Stenholm:1986}. In this limit it is convenient to use the Wigner function $f_{\rm tot}({\bf x},{\bf p},t)$ for the atomic variables ${\bf x}=(x_1,\ldots,x_N)$, ${\bf p}=(p_1,\ldots,p_N)$:
\begin{equation}
f_{\rm tot}({\bf x},{\bf p},t)=\int\frac{d^Ny}{(2\pi\hbar)^N} \,{\rm e}^{-\frac{i}{\hbar}{\bf y}\cdot {\bf p}}\,\mathrm{Tr}\left\{\ket{{\bf x}-\frac{1}{2}{\bf y}}\bra{{\bf x}+\frac{1}{2}{\bf y}}\hat{\rho}(t)\right\}\,,
\end{equation}
with ${\bf y}=(y_1,\ldots,y_N)$. We further assume that the cavity field relaxes very fast to a local steady state depending on the atomic distribution, which is verified when the inequality $k\Delta p/m\ll|\kappa_n+i\Delta_n|$ is fulfilled, namely, when the dimensionless parameter 
\begin{equation}
\varepsilon=\frac{k\Delta p/m}{|\kappa_n+i\Delta_n|} \label{eq:epsilon}
\end{equation}  
is small. This allows us to identify a coarse-grained time scale $\Delta t$ that is infinitesimal for the external degrees of freedom but over which the cavity degrees of freedom can be eliminated from the equations of the atomic dynamics. In particular, $f_{\rm tot}({\bf x},{\bf p},t)=f({\bf x},{\bf p},t)+f_{\rm na}({\bf x},{\bf p},t)$, where $f({\bf x},{\bf p},t)$ is the term in zero order in the retardation effect, corresponding to the cavity field following adiabatically the atomic motion, and $f_{\rm na}({\bf x},{\bf p},t)$ represents the non-adiabatic corrections scaling with $\varepsilon$. The latter can be expressed in terms of $f({\bf x},{\bf p},t)$ using perturbation theory \cite{Dalibard:1985,Schuetz:2013}. The derivation is lengthy but is also a straightforward extension of the derivation for a single-mode cavity, which is extensively reported in Ref. \cite{Schuetz:2013}. We thus refer the interested reader to this work and present here the resulting FPE, which reads 
\begin{eqnarray}
&\frac{\partial}{\partial t}f({\bf x},{\bf p},t)+\left\lbrace f({\bf x},{\bf p},t),H({\bf x},{\bf p})\right\rbrace =\nonumber\\ 
&\sum_{i,j=1}^N\sum_{n=1,2}\frac{\partial}{\partial p_i}\left[\sin(k_nx_i)\sin(k_nx_j)\left(D_n\frac{\partial f({\bf x},{\bf p},t)}{\partial p_j}-\Gamma_n p_j f({\bf x},{\bf p},t)\right)\right] \nonumber\\
&+\sum_{i,j=1}^N\sum_{n=1,2}\frac{\partial}{\partial p_j}\left[\eta_n\sin(k_nx_i)\sin(k_nx_j)\frac{\partial}{\partial x_i}f({\bf x},{\bf p},t)\right]\,.
\label{eq:fpe}
\end{eqnarray}
In detail, Hamiltonian (\ref{eq:hamiltonian}) results from the adiabatic component of the dynamics and describes coherent long-range, two-body interactions which are mediated by the cavity photons:
\begin{equation}
H({\bf x},{\bf p})=\sum_{i=1}^N\frac{p_i^2}{2m}-N\sum_{n=1,2}\gamma_n\Theta_n^2\,,
\label{eq:hamiltonian}
\end{equation}
where $\gamma_n$ is a scalar and 
\begin{equation}
\Theta_1=\frac{1}{N}\sum_{i=1}^N\cos(kx_i/2)\,; \, \Theta_2=\frac{1}{N}\sum_{i=1}^N\cos(kx_i)\,.
\end{equation}
The quantities $\Theta_n$ are order parameters for spatial selforganization. In fact, they vanish for homogeneous spatial distributions, they are both different from zero when the atomic density forms spatial gratings with periodicity $4\pi/k$, while $\Theta_1=0$ and $\Theta_2\neq 0$ when the spatial grating has periodicity $2\pi/k$, as illustrated in Fig. \ref{Fig:2}. The cavity mode field amplitudes $\mathcal E_{n}=\langle\hat{a}_n\rangle$, in turn, are proportional to $\Theta_n$ and in leading order in the expansion in $\varepsilon$ read \cite{Schuetz:2015}
 \begin{equation}
\mathcal E_{n}=\frac{NS_n\Theta_n}{\Delta_n+i\kappa_n}\,.
\label{eq:steady_cavity_field}
\end{equation}
The Bragg gratings can be stable provided that $\gamma_n>0$. The sign of $\gamma_n$ is here controlled by the detuning $\Delta_n$, and hence by the frequency of the pumping laser. For later convenience we write $\gamma_n=\alpha_n/\beta_n$ with 
\begin{eqnarray}
&&\alpha_n=\frac{4NS_n^2\Delta_n^2}{(\Delta_n^2+\kappa_n^2)^2}\,,\label{alphan}\\
&&\beta_{n}=\frac{-4\Delta_n}{\hbar(\Delta_n^2+\kappa_n^2)}\,. \label{betan}
\end{eqnarray}
Friction and diffusion are instead due to retardation effects between atoms and cavity dynamics and describe cross-correlations between the atoms, which can play a relevant role in stabilizing the system in non-thermal metastable states \cite{Schuetz:2016}. Their coefficients take the form
\begin{eqnarray}
D_n&=(\hbar k_n)^2 S_n^2\frac{\kappa_n}{\Delta_n^2+\kappa_n^2}\,,\\
\Gamma_n&=\frac{\hbar k_n^2}{m}S_n^2\frac{4\Delta_n\kappa_n}{(\Delta_n^2+\kappa_n^2)^2}\,,\\
\eta_n&=\frac{(\hbar k_n)^2}{m}S_n^2\frac{\kappa_n^2-\Delta_n^2}{(\Delta_n^2+\kappa_n^2)^2}\,,
\end{eqnarray}
and are here reported in the limit $|\Delta_n|\gg NU_n$, where we neglected the contribution of the dynamical Stark shift to the dynamics. 

\subsection{Existence of a stationary state}

The FPE  (\ref{eq:fpe}) allows for a stationary solution satisfying the condition $\partial_t  f_{\mathrm{st}}({\bf x},{\bf p},t)=0$. We first consider two limiting cases, in which only one cavity mode is pumped. These situations correspond to the dynamics of atoms in a single-mode standing-wave cavity investigated in Refs. \cite{Schuetz:2014,Schuetz:2015, Schuetz:2013,Jaeger:2016}. 

Let us first assume that $S_1=0$ but $S_2\neq 0$. In this case a stationary state exists provided that $\Delta_2<0$ and the stationary distribution reads $f_{\mathrm{st}}=C_2\exp(-\beta_2 H|_{\gamma_1=0})$, with normalization constant $C_2$ and $\beta_2=-\frac{m\Gamma_2}{D_2}$ given in Eq. (\ref{betan}) \footnote{In this discussion we neglect the terms of Eq. (\ref{eq:fpe}) that scale with $\eta_n$. This is exact if $\Delta_n=-\kappa_n$. In general, these terms give rise to corrections to the coherent dynamics that scale with $1/N$ in the thermodynamic limit we use in Sec. \ref{Sec:3}, see Ref. \cite{Jaeger:2016} for details.}. Vice versa, when $S_1\neq 0$, $S_2=0$, and $\Delta_1<0$, the stationary state is $f_{\mathrm{st}}=C_1\exp(-\beta_1 H|_{\gamma_2=0})$ with $\beta_1=-\frac{m\Gamma_1}{D_1}$ from Eq. (\ref{betan}). 

When both resonators are pumped, a stationary solution can be found provided that $\Delta_1,\Delta_2$ are negative and $\beta_1=\beta_2$, namely:
\begin{equation}
\frac{\Delta_1}{\Delta_1^2+\kappa_1^2}=\frac{\Delta_2}{\Delta_2^2+\kappa_2^2}\,. \label{condition}
\end{equation}
This is the situation we consider in the following. In particular, $\beta_1=\beta_2=\beta$, where
 \begin{equation}
\beta^{-1}=\frac{\hbar(\Delta_n^2+\kappa_n^2)}{-4\Delta_n }\,. \label{beta0}
\end{equation}
The detunings and cavity loss rates thus determine an effective temperature $T_{\rm eff}=k_B/\beta$ characterizing the stationary state. The stationary state is given by
\begin{equation}
f_{\mathrm{st}}=C\exp(-\beta H)\,.\label{eq:steady}
\end{equation}

\section{Mapping to the generalized Hamiltonian Mean Field model} \label{Sec:3}

We now consider the stationary state of Eq. (\ref{eq:steady}) and draw a formal analogy to a canonical ensemble at equilibrium.  For this purpose we define the thermodynamic limit, according to which $\alpha_n\propto NS_n^2$ is constant. This assumption warrants that the energy is extensive, it is thus equivalent to Kac scaling \cite{Campa:2009} and physically corresponds to scaling the cavity mode volume linearly with the number of particles \cite{Asboth:2005,Fernandez-Vidal:2010}. In this thermodynamic limit we obtain an explicit expression of the free energy per particle which allows us to perform a mapping of the steady state in Eq. (\ref{eq:steady}) to the canonical ensemble realization of the GHMF.  

We determine the free energy per particle $\mathcal{F}$ using the relation $\mathcal{F}=-\ln(Z)/(N\beta)$, where $Z$ is the ``canonical" partition function and reads:
\begin{equation}
Z=\frac{1}{h^N}\int_0^{\lambda} dx_1\ldots\int_0^{\lambda} dx_N\int_{-\infty}^{\infty} dp_1\ldots\int_{-\infty}^{\infty} dp_Ne^{-\beta H(\textbf{x},\textbf{p})} ,
\end{equation}
with $\lambda=4\pi/k$. We integrate over the momenta $\textbf{p}$ and apply the Hubbard-Stratonovich transformation to eliminate $\Theta_n^2(\textbf{x})$, obtaining
\begin{eqnarray}
Z= \frac{N\sqrt{\alpha_1\alpha_2}}{\pi\left(\pi\hbar\omega_r\beta\right)^{\frac{N}{2}}}\int_{-\infty}^{\infty}dy_1\int_{-\infty}^{\infty}dy_2\exp\left[-N\left\lbrace\alpha_1 y_1^2+\alpha_2 y_2^2-\ln\left(\mathcal{I}(y_1,y_2)\right)\right\rbrace\right]\nonumber \,,
\end{eqnarray} 
with $q_i=kx_i/2$, $\omega_r=\hbar k^2/(2m)$, and 
\begin{equation}
\mathcal{I}(y_1,y_2)=\int_{0}^{2\pi}dq\exp\left[2\left(\alpha_1 y_1\cos(q)+\alpha_2 y_2\cos(2q)\right)\right]\,.
\label{eq:canonical_integral}
\end{equation}
In the thermodynamic limit $N\rightarrow\infty$ we perform a saddle-point approximation, which leads to the expression for the free energy
\begin{equation}
\mathcal{F}(y_1,y_2)=\frac{\ln\left(\pi\hbar\omega_{\mathrm{r}}\beta\right)}{2\beta}+\frac{1}{\beta}\inf_{y_1,y_2}\left\lbrace\alpha_1 y_1^2+\alpha_2 y_2^2-\ln\left(\mathcal{I}(y_1,y_2)\right)\right\rbrace
\label{eq:free_energy}\,.
\end{equation}
This expression coincides, apart for irrelevant constants, with the mean free energy of the GHMF model in the canonical ensemble \cite{Pikovsky:2014}. In particular, the extrema $y_n^*$ of the free energy fulfill the relation
\begin{equation}
y_n^*=\frac{\int_{0}^{2\pi}dq\cos(nq)\exp\left[2(\alpha_1y_1^*\cos(q)+\alpha_2y_2^*\cos(2q))\right]}{\int_{0}^{2\pi}dq \exp\left[2(\alpha_1y_1^*\cos(q)+\alpha_2y_2^*\cos(2q))\right]}\,, \label{eq:fixpoint}
\end{equation}
their values lie in the interval $y_n^*\in [-1,1]$ and they can be identified with the variables $\Theta_n$: $y_n^*=\langle \cos nq\rangle $. This shows that $\Theta_1,\Theta_2$ are analogous to the magnetization in the GHMF model. By means of this mapping, moreover, we connect the lasers and cavity parameters with the dimensionless parameter $\Delta$ of the GHMF in Eq. (\ref{eq:GHMF}): $\alpha_1/\beta\to \Delta$ and $\alpha_2/\beta \to (1-\Delta)$. Therefore, varying the lasers' amplitudes would allow one to span over the values of $\Delta$ in Eq. (\ref{GHMF}), while the effective temperature can be tuned varying the detunings $\Delta_1$ and $\Delta_2$, and thus the lasers' frequencies (provided condition (\ref{condition}) is fulfilled). 

\begin{figure}
\centering
\includegraphics[width=0.8\textwidth]{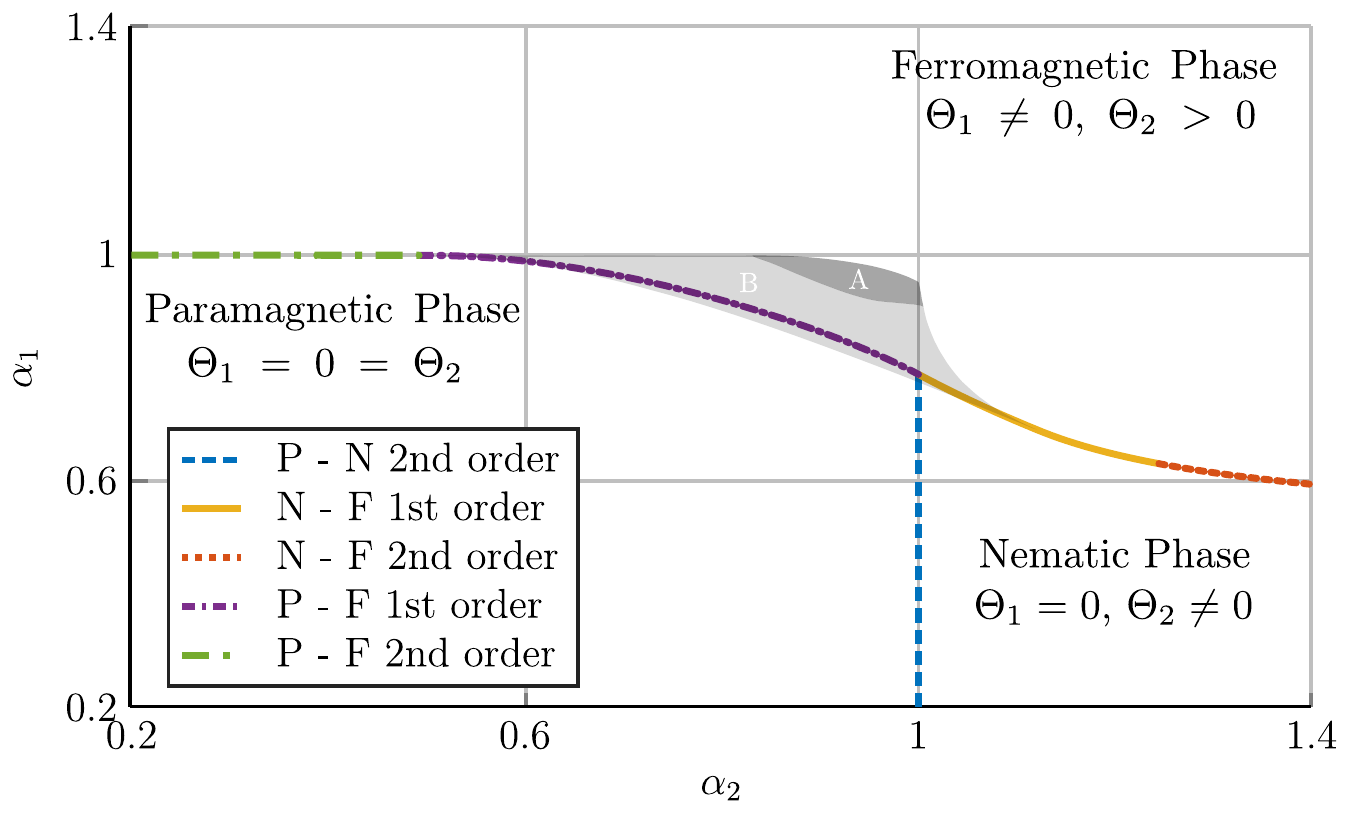}
\caption{Canonical phase diagram as a function of $\alpha_1$ and $\alpha_2$. The phases are identified by numerically determining the global minima of the free energy of Eq. (\ref{eq:free_energy}). The system shows second-order and first-order phase transitions (see inset). In the inset: P denotes paramagnetic, F ferromagnetic, and N nematic phase. The dark and light gray areas, labeled by A and B, respectively, indicate the parameter regions where ensemble inequivalence is expected. In A (B) the microcanonical ensemble exhibits three (two) phases \cite{Pikovsky:2014}.}
\label{Fig:3}
\end{figure}

The phase diagram is obtained following the same analysis as in Ref. \cite{Pikovsky:2014} and it is illustrated in Fig. \ref{Fig:3}. The system exhibits second and first order phase transitions (see \ref{App:A} for details), which are found for the same corresponding values of the phase diagram as in  Ref. \cite{Pikovsky:2014}. The phases can be measured by monitoring the amplitude and the phase of the fields at the cavity output, since these are proportional to the order parameters $\Theta_1$ and $\Theta_2$, as visible in Eq. (\ref{eq:steady_cavity_field}). 

\section{Discussion and outlook} \label{Sec:4}

The mapping  of the stationary dynamics of Eq. (\ref{eq:fpe}) to the canonical GHMF shows that cavity QED can be a versatile platform for studying equilibrium statistical mechanics of long-range interacting systems. In this perspective it is important to identify the parameter regimes for which this setup could simulate the microcanonical GHMF model. This would allow one to experimentally investigate the ensemble inequivalence that has been predicted for the GHMF \cite{Pikovsky:2014}.

Within the validity of the semiclassical description, the microcanonical GHMF could be realised in the regime where the parameter $\varepsilon$, Eq. (\ref{eq:epsilon}), becomes smaller than the small parameter $\hbar k/\Delta p$ of the semiclassical expansion. This would require one to choose the detunings $|\Delta_n|\gg\kappa_n$ \cite{Fernandez-Vidal:2010}. In this limit there is a well defined time scale over which the dynamics is coherent and solely dominated by Hamiltonian (\ref{eq:hamiltonian}), while the right-hand side of the FPE (\ref{eq:fpe}), which scales with $\varepsilon$, can be discarded.  Moreover, in order to prepare the system in a microcanonical ensemble, the atomic gas 
shall be in the asymptotic Boltzmann-Gibbs distribution of the corresponding Hamiltonian dynamics \cite{Campa:2009,Levin:2014}. By identifying the mean energy per particle in the canonical ensemble $\langle E/N\rangle=\partial(\beta\mathcal{F})/\partial\beta$ with the constant energy $\epsilon$ in the microcanonical ensemble, we obtain the relation
\begin{equation}
\epsilon=\frac{1}{2\beta}-\frac{1}{\beta}\left(\alpha_1\Theta_1^2+\alpha_2\Theta_2^2\right),\label{energy}
\end{equation}
which allows us to directly compare the results obtained in both ensembles. Ensemble inequivalence is predicted in the regions where the three phases meet and is indicated by the shaded area in Fig. \ref{Fig:3}. In an experimental realization it would become evident by detecting different phases in the canonical and microcanonical realizations for the same values of $\alpha_1$ and $\alpha_2$ \cite{Pikovsky:2014}. It would be interesting to identify observables of this system which provide a measure of the specific heat, thus allowing one to determine whether it may become negative in the microcanonical ensemble. This is an open question which we will address in future work.

The authors acknowledge discussions with Shamik Gupta, Julian L\'eonard, Stefan Sch\"utz, Valentin Torggler, and Tobias Donner. This work was supported by the German Research Foundation (DFG DACH "Quantum crystals of matter and light") and by the European Commission (ITN network "ColOpt"). 

\begin{appendix}

\section{Determination of the canonical phase diagram} \label{App:A}

\begin{figure}[h!]
		\centering
		\includegraphics[width=0.8\textwidth]{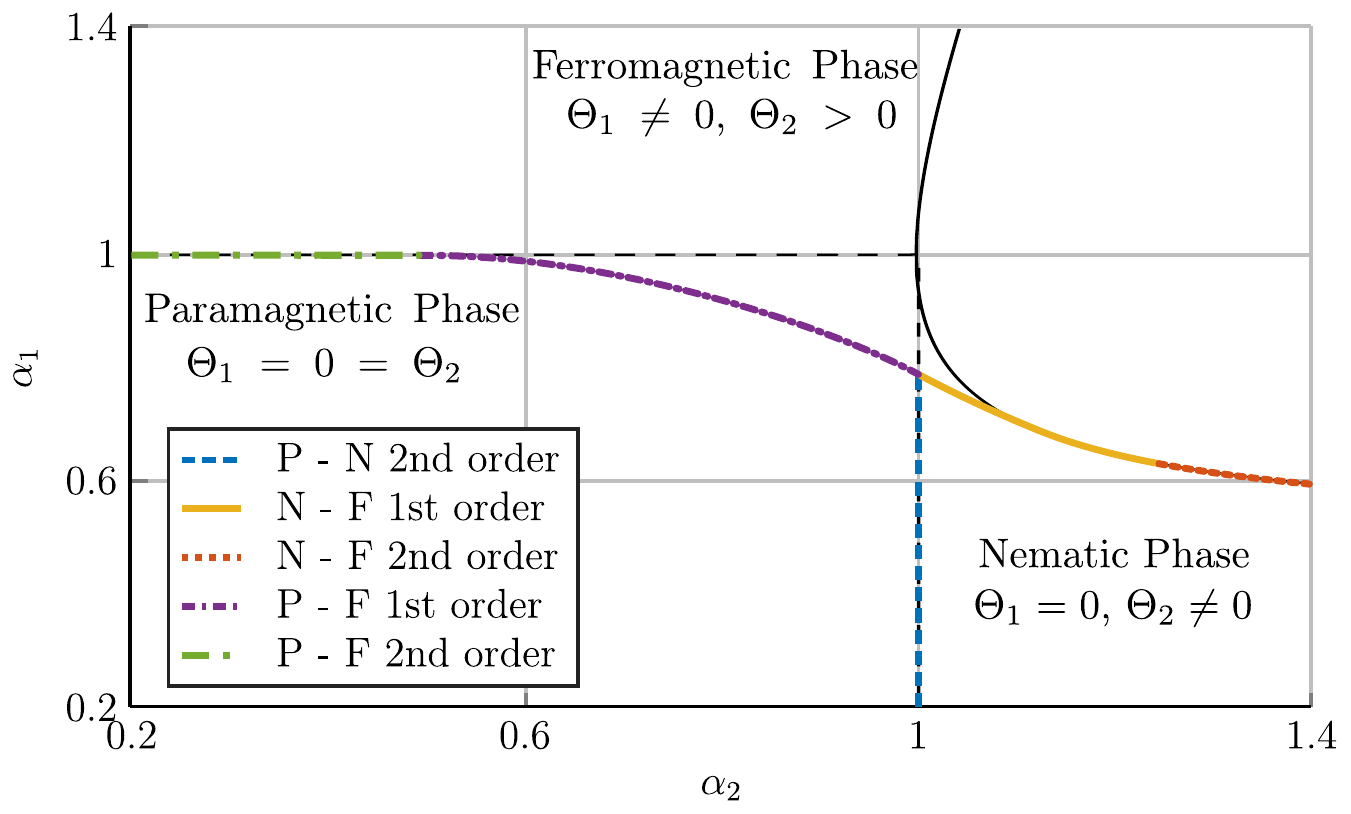}
		\caption{The canonical phase diagram as in Fig. \ref{Fig:3} and the results of our analytical analysis. The dashed black lines delimitate the region determined by the inequality Eq.(\ref{eq:canonical_para_conditions}), where a paramagnetic phase ($\Theta_1=0=\Theta_2$) is a local minimum of the free energy. The area below the solid black line (determined by the inequality Eq.(\ref{eq:canonical_nematic_conditions1}) ) is the region where a nematic phase  ($\Theta_1=0$, $\Theta_2\neq 0$) is a local minimum of the free energy.}
		\label{Fig:4}
	\end{figure}

For the calculation of the phase diagram in Fig. \ref{Fig:4} we numerically calculated the global minimum of the free energy, Eq. (\ref{eq:free_energy}), for each pair $\boldsymbol{\alpha}=(\alpha_1,\alpha_2)$. The global minimum describes a paramagnetic phase when $\Theta_1(\boldsymbol{\alpha})=\Theta_2(\boldsymbol{\alpha})=0$. The phase is nematic when $\Theta_1(\boldsymbol{\alpha})=0$ and $\Theta_2(\boldsymbol{\alpha})\neq 0$ and ferromagnetic when $\Theta_1,\Theta_2\neq 0$. A phase transition occurs where the properties of the minimum change by varying  $\boldsymbol{\alpha}$. The order of the transition is determined by calculating numerically the first derivatives of $ \mathcal{F}(\boldsymbol{\alpha})$ with respect to $\alpha_1$ and $\alpha_2$: If they are discontinuous at the phase transition the transition is of first order, while if they are continuous the transition is of second order. This determines the phase diagram in Fig. \ref{Fig:4}.\\

We also checked our results by analytically calculating the Hessian matrix of the free energy, Eq. (\ref{eq:free_energy}), at the extrema where Eq. (\ref{eq:fixpoint}) is fulfilled. We note that $\Theta_1=\Theta_2=0$ is always a solution of Eq. (\ref{eq:fixpoint}). It is a minimum when the following inequalities hold:
\begin{equation}
	\alpha_1<1\,, \qquad\alpha_2<1\,.
	\label{eq:canonical_para_conditions}
	\end{equation}
The thin black dashed line in Fig. \ref{Fig:4} delimitates the region $\alpha_1,\alpha_2<1$, where the paramagnetic phase is a local minimum of the free energy.
We now consider the nematic phase, set $\Theta_1=0$ in Eq. (\ref{eq:fixpoint}) and obtain the equation for $\Theta_2$:
\begin{equation}
\Theta_2=\frac{I_1\left(2\alpha_2\Theta_2\right)}{I_0\left(2\alpha_2\Theta_2\right)}\,,\label{Besselfunction}
\end{equation}
where $I_n(x)$ is the modified Bessel function of order $n$. The Hessian matrix is positive definite when the following inequalities are fulfilled:
\begin{eqnarray}
	&\alpha_1<\frac{1}{1+\Theta_2}\,,	\label{eq:canonical_nematic_conditions1} \\
	1<&\alpha_2<\frac{1}{1-\Theta_2^2}\,.
	\label{eq:canonical_nematic_conditions2}
\end{eqnarray}	
One can show that imposing $\Theta_2\neq 0$ in Eq. (\ref{Besselfunction}) is equivalent to Eq. (\ref{eq:canonical_nematic_conditions2}). Inequality (\ref{eq:canonical_nematic_conditions1}) determines an upper threshold $\alpha_{1,c}=1/(1+\Theta_2)$ on $\alpha_1$ above which the nematic configuration is no longer a minimum of the free energy. The thin black solid line in Fig. \ref{Fig:4} shows $\alpha_{1,c}$ in the $\boldsymbol{\alpha}$-plane. Note that $\alpha_{1,c}<1$ ($\alpha_{1,c}>1$) for $\Theta_2>0$ ($\Theta_2<0$). In the limit $\alpha_2\to\infty$, Eq. (\ref{Besselfunction}) delivers  $\Theta_2\to 1$ and $\Theta_2\to-1$, giving $\alpha_{1,c}\to1/2$ and $\alpha_{1,c}\to\infty$, respectively. \\
We note that the conditions we determine analytically do not overlap for all values of $\alpha_1,\alpha_2$ with the numerically calculated phase transition lines (see Fig. \ref{Fig:4}).
For instance there is an area where the inequalities (\ref{eq:canonical_para_conditions}) hold but the global minimum is in the ferromagnetic phase. Moreover the condition in Eq. (\ref{eq:canonical_nematic_conditions1}) predicts a parameter region where the nematic phase with $\Theta_2<0$ is a local minimum but the global minimum is a ferromagnetic phase. 
\end{appendix}

\appendix

\end{document}